\documentclass[prl, superscriptaddress,twocolumn,10pt
 amsmath,
 amssymb,
 aps,
]{revtex4}

\usepackage{graphicx}
\usepackage{dcolumn}
\usepackage{bm}
\usepackage{physics}
\usepackage{color}
\usepackage{float}

\definecolor{darkpink}{rgb}{0.91, 0.33, 0.5}

\begin{document}

\preprint{APS/123-QED}

\title{High-harmonic generation in spin-orbit coupled systems}

%
%
%

\author{Markus Lysne}
\affiliation{Department of Physics, University of Fribourg, 1700 Fribourg, Switzerland}
\author{Yuta Murakami}
\affiliation{Department of Physics, Tokyo Institute of Technology, Meguro, Tokyo 152-8551, Japan}
\author{Michael Sch\"uler}
\affiliation{Stanford Institute for Materials and Energy Sciences (SIMES),
SLAC National Accelerator Laboratory, Menlo Park, CA 94025, USA}
\author{Philipp Werner}
\affiliation{Department of Physics, University of Fribourg, 1700 Fribourg, Switzerland}

\date{\today}

\begin{abstract}
We study high-harmonic generation in two-dimensional electron systems with Rashba and Dresselhaus spin-orbit coupling and derive harmonic generation selection rules with the help of group theory. Based on the bandstructures of these minimal models and explicit simulations we reveal how the spin-orbit parameters control the cutoff energy in the high-harmonic spectrum.  We also show that  
the magnetic field and polarization dependence of this spectrum provides information on the magnitude of the Rashba and Dresselhaus spin-orbit coupling parameters. The shape of the Fermi surface can be deduced at least qualitatively and if only one type of spin-orbit coupling is present, the coupling strength can be determined.

\end{abstract}

\maketitle

{\it Introduction.}
The concept of high harmonic generation (HHG) has attracted interest for decades in atomic systems, and lately also in the condensed matter community \cite{Lewenstein_1994, Ghimire_2018, Krausz_2009, Ghimire_2011}. It is a non-linear process in which a system driven by light at a certain frequency can give rise to emission at multiples of this fundamental frequency \cite{Vampa_2014, Luu_2015}. In the condensed matter context, one of the interesting aspects is that the harmonic spectrum carries information about the microscopic model, like the band structure or interaction parameters \cite{Luu_2015,Vampa_2015, Murakami_2018a, Lysne_2020}.

Several mechanisms have been proposed to explain the many facets of HHG. In some cases, the physics can be understood within a single-particle picture, but still necessitates the numerical evaluation of the interband polarization and intraband current in a coupled set of equations \cite{Golde_2008, Vampa_2014,Hohenleutner_2015}. Which of these two processes dominates the emission has been debated for a long time,  
and 
a unified HHG mechanism applicable to a wide range of solids is still lacking \cite{Liu_2017}. Recently, 
the scope of HHG studies has been extended to  
strongly correlated systems \cite{Silva_2018, Murakami_2018a, Murakami_2018b, Imai_2020, Dejean_2018, Lysne_2020}, disordered systems \cite{Yu2019PRB2,Almalki2018PRB,Orlando2018,Ikeda2019}, the effects of spin-polarized defects \cite{Mrudul_2020}, spin or multiferroic systems \cite{Takayoshi_2019,ikeda2019high}, HHG in graphene and transition metal dicalchogenides \cite{Yoshikawa_2017, Yoshikawa_2019}, to mention a few.  

The effect of spin-orbit coupling (SOC) on HHG has to the best of our knowledge not been systematically explored. SOC is a relativistic effect in solids which locks the spin direction in relation to the electron momentum \cite{Manchon_2015, Hirohata_2020, Igor_2004}. It acts on the electron's motion like an effective momentum-dependent magnetic field 
and gives rise to an {\it intrinsic} spin Hall effect \cite{Shen_2004, Manchon_2015}. 
SOC plays an important role in topological insulators (TI) \cite{Bernevig_2013} and HHG has been used as a tool for detecting topological properties such as the Berry curvature \cite{Luu_2018}. In this letter, we however want to study the effect of SOC in an isolated way, focusing on minimal models of two-dimensional (2D) SOC systems.
The goal is to understand how SOC affects the HHG cutoffs and how the type of SOC and the coupling parameters can be extracted from characteristic features of the spectrum. For our analysis, we will adapt the existing theory for harmonic generation (HG) selection rules to models defined in momentum space \cite{PhysRevLett.80.3743, Neufeld_2019}, using concepts similar to  
non-symmorphic symmetries in Floquet topological insulators \cite{Morimoto_2017}. While this type of symmetry analysis has been used before \cite{Saito_2017, Neufeld_2019, Ikeda_2018},  
it is formulated here in a way which is convenient for SOC systems.

{\it Model and symmetries.}
Several previous works have discussed the Rashba and Dresselhaus Hamiltonians in a tight-binding framework \cite{PhysRevB.64.024426, pareek2002spin, pareek2001magnetic}, which yields
\begin{align}
    &\hat{H} = \sum_{\boldsymbol{k}}  \Psi_{\boldsymbol{k}}^{\dagger} [\epsilon(\boldsymbol{k})\otimes \sigma_0 -(\alpha \sin(k_y a) -\gamma \sin(k_x a))\otimes \sigma_x \nonumber\\
    &\quad \quad +(\alpha \sin(k_x a) -\gamma \sin(k_y a)) \otimes \sigma_y + B \sigma_z] \Psi_{\boldsymbol{k}}, 
    \label{momHam}
\end{align}
with $\Psi_{\boldsymbol{k}} = (\hat{c}_{\boldsymbol{k},\uparrow}, \hat{c}_{\boldsymbol{k},\downarrow})^T$ a spinor combining the annihilation operators for momentum $\boldsymbol{k}$ and spin up and down, $\epsilon(\boldsymbol{k})=2t_h(4-\cos(k_x a) - \cos(k_y a))$ the dispersion of the lattice, $\sigma_{0}$ the identity matrix and $\sigma_{i=x,y,z}$ the Pauli matrices. $\alpha$ denotes the strength of the Rashba SOC, $\gamma$ that of the Dresselhaus SOC \cite{rashba2006spin}, $B$ an external magnetic field and/or exchange field \cite{Ado_2016, Qaiumzadeh_2015}, which is assumed to couple only to the spin (no Landau levels). $a$ and $t_h$ are the lattice spacing and the hopping parameter, respectively,
and we will set both to $1$ in the following.
 This type of SOC represents the most typical form in 2D materials \cite{Winkler_2003}. We incorporate the electric field through the Peierls substitution $\boldsymbol{k} \rightarrow \boldsymbol{k} + \boldsymbol{A}(t)$, where $\boldsymbol{A}(t)$ denotes the vector potential. When developing selection rules for the HHG spectra, we will assume an AC field driving with frequency $\Omega=\tfrac{2\pi}{T}$, so that the Hamiltonian satisfies $\hat{H}(t)=\hat{H}(t+T)$. Following Refs.~\onlinecite{PhysRevLett.80.3743} and \onlinecite{Neufeld_2019} we 
combine operators acting on space, time and spin to define symmetry operations for the periodically driven system. This analysis is applicable to Hamiltonians of the general form $\hat{H}(t) = \sum_{\boldsymbol{k}} \Psi_{\boldsymbol{k}}^{\dagger} h(\boldsymbol{k},t) \Psi_{\boldsymbol{k}}$.

The process of identifying the HG selection rules for a momentum resolved quantity $\mathcal{O}(\boldsymbol{k},t)$ contains the following two steps:
(i) 
Identify a group, $G$, of symmetry operations $g$, leaving $h(\boldsymbol{k},t)$ invariant, i.e., $gh(\boldsymbol{k},t)g^{-1}=h(\boldsymbol{k},t)$. 
(ii)	
	For $g\in G$ analyze the restrictions on $n$ which follow from the condition $\hat{g} \mathcal{O}(\boldsymbol{k},t) e^{in\Omega t} \hat{g}^{-1} \equiv \mathcal{O}(\boldsymbol{k},t) e^{in\Omega t}$ for the generator $g$ of $G$. The latter requirement is equivalent to saying that $\mathcal{O}(\boldsymbol{k},t) e^{in\Omega t}$ belongs to the trivial representation of $G$ \cite{PhysRevLett.80.3743}. 

As the density matrix is also a time dependent quantity which must be factored into the calculation of any observable, we assume that 
(ii) holds for the density matrix \emph{and} the observable combined, i.e., $\hat{g} \rho(t) \mathcal{O}(\boldsymbol{k},t) e^{in\Omega t} \hat{g}^{-1} = \rho(t) \mathcal{O}(\boldsymbol{k},t) e^{in\Omega t}$.  
Returning to model (\ref{momHam}) we begin by listing the generators of symmetry groups which are isomorphic to some cyclic group $\boldsymbol{Z}_n$. 
One symmetry which holds for both linearly and circularly polarized light is
\begin{equation}
  \mathcal{P} \otimes \mathcal{T}_2 \otimes \mathcal{S}_{(-)} =\begin{cases}
	&\boldsymbol{k} \rightarrow -\boldsymbol{k} \nonumber\\
	&t \rightarrow t + T/2 \\
	& (\sigma_x, \sigma_y, \sigma_z) \rightarrow (-\sigma_x,-\sigma_y, \sigma_z) \nonumber
  \end{cases}.
\end{equation} 
For the Rashba-Dresselhaus model with $\alpha=\pm\gamma$, and for $B=0$, we have the additional symmetry 
\begin{equation}
  \mathcal{P} \otimes \mathcal{T}_2 \otimes \mathcal{S}_{x,\mp y} =\begin{cases}
	&\boldsymbol{k} \rightarrow -\boldsymbol{k} \nonumber\\
	&t \rightarrow t + T/2 \\
	& \sigma_{x(y)} \rightarrow \mp \sigma_{y(x)} \nonumber
  \end{cases},
\end{equation}
also valid for circular and linear polarization. In the case of circularly polarized light, where we define $\boldsymbol{A}(t) = (A_x(t),A_y(t))=A_0(\sin(\Omega t),\cos(\Omega t))$, the following additional symmetries are found
\begin{equation} 
  \mathcal{R}_{90^{\circ}} \otimes \mathcal{T}_4 \otimes \mathcal{S}_{\pm90^{\circ}} =\begin{cases}
	& (k_x,k_y) \rightarrow (k_y,-k_x)  \\
	&t \rightarrow t + T/4 \\
	&  (\sigma_x, \sigma_y, \sigma_z) \rightarrow (\pm\sigma_y, \mp\sigma_x, \sigma_z) 
  \end{cases},
  \label{circSym}
\end{equation}
where the upper sign is for $\gamma=0$ and the lower sign for $\alpha=0$. 
Note the $t \rightarrow t + T/4$ transforms $\boldsymbol{A}$ as 
$(A_x,A_y) \rightarrow (A_y,-A_x)$.
For the case where both $\alpha \neq 0$ and $\gamma \neq 0$, there is no symmetry involving $t \rightarrow t + T/4$, because the Fermi surfaces only have a two-fold rotational symmetry \cite{Manchon_2015}.  

{\it Selection rules.}
For linearly polarized light described by $A_x(t) = A_0 \cos(\Omega t)$, let us consider the charge velocity 
\begin{align}
	&v_x(\boldsymbol{k},t) =\frac{\partial h(\boldsymbol{k},t)}{\partial k_x} = 
	2\sin(k_x + A_x(t)) \cdot \sigma_0 \nonumber\\
	&\hspace{2mm}+ \alpha \cos(k_x + A_x(t)) \cdot \sigma_y 
	 + \gamma \cos(k_x + A_x(t)) \cdot \sigma_x 
	\label{chargeVelR}
\end{align}
and the symmetry ${g}\equiv\mathcal{P} \otimes \mathcal{T}_2 \otimes \mathcal{S}_{(-)}$. Since the spin transformation does not mix the $\sigma_0$, $\sigma_x$ and $\sigma_y$ matrices it is sufficient to consider $\hat{g}\cos(k_x + A_x(t))e^{in \Omega t} \sigma_{x,y}\hat{g}^{-1}=\cos(-k_x - A_x(t))e^{in \Omega (t+T/2)} (-\sigma_{x,y}) \stackrel{!}=\cos(k_x + A_x(t))$ $\times e^{in \Omega t} \sigma_{x,y}$, which leads to $e^{in\pi}=-1$, i.e., {\it odd} order harmonics only. 
We proceed to demonstrate how the spin current - a quantity often studied in spintronics applications - is linked to HHG \cite{Igor_2004}. The momentum-resolved spin current operator is defined as \cite{Rashba_2003} 
\begin{equation} \label{spinCurrDef}
	\mathcal{V}_{ij}(\boldsymbol{k},t) = \tfrac12 \big( \sigma_i\cdot v_{j}(\boldsymbol{k},t)  +  v_{j}(\boldsymbol{k},t)\cdot\sigma_i\big),
\end{equation}
where $i,j=x,y,z$. We denote the spin current by $J_{ij}(t) = N_{\boldsymbol{k}}^{-1}\sum_{\boldsymbol{k}} \textrm{Tr}[\rho(t) \mathcal{V}_{ij}(\boldsymbol{k},t)]$ with $N_{\boldsymbol{k}}$ the number of ${\boldsymbol k}$ points in the Brillouin zone. Using the group generator 
$\hat{g} \equiv  \mathcal{P} \otimes \mathcal{T}_2 \otimes \mathcal{S}_{(-)}$ on 
\begin{equation} \label{spinVel}
	\mathcal{V}_{yx}(\boldsymbol{k},t) = \gamma \cos(k_x + A_x(t)) \cdot \sigma_0
	+2 \sin(k_x + A_x(t)) \cdot \sigma_y,
\end{equation}
the constraint following from $\hat{g} \mathcal{V}_{yx}(\boldsymbol{k},t)  e^{in\Omega t}\hat{g}^{-1} \equiv \mathcal{V}_{yx}(\boldsymbol{k},t) e^{in\Omega t}$ implies {\it even} order harmonics only. (The presence of a Dresselhaus SOC yields a non-zero $\mathcal{V}_{xx}$ component, which gives the same constraint on the harmonics.) This prediction is consistent with Ref.~\onlinecite{Hamamoto_2017}, which discussed a second harmonic signal in the spin current based on symmetry arguments. 
Furthermore, one can apply selection rules to $\mathcal{V}_{zy}$ which is the component relevant for the spin Hall effect \cite{Shen_2004}. As this expression is the anti-commutator of a term with even harmonics 
and one with odd harmonic orders, the result is \emph{odd}. 

Let us briefly mention the role of inversion symmetry. We have seen that despite the breaking of inversion symmetry (which has been linked to even order harmonics \cite{Manchon_2015}) model (\ref{momHam}) only produces odd order harmonics in the longitudinal charge velocity.  
To understand this let us consider the general Hamiltonian $\hat{H}(t)=\sum_{\boldsymbol{k}} \hat{c}_{\boldsymbol{k},\alpha}^{\dagger} h_{\alpha,\beta}(\boldsymbol{k} + \boldsymbol{A}(t)) \hat{c}_{\boldsymbol{k},\beta}$ with the indices denoting the relevant (orbital, spin etc.) degrees of freedom. In the Rashba model, we clearly have that $h_{\alpha,\beta}(\boldsymbol{k}+\boldsymbol{A}(t)) \neq h_{\alpha,\beta}(-\boldsymbol{k}-\boldsymbol{A}(t))$. However, as long as there exists a transformation $g$ for which the action on the additional degrees of freedom yields $h_{\alpha,\beta}(\boldsymbol{k} + \boldsymbol{A}(t)) = h_{g(\alpha),g(\beta)}(-\boldsymbol{k}- \boldsymbol{A}(t))$, the HHG radiation is restricted to odd harmonics. 

To test the selection rules, we simulate model \eqref{momHam} at inverse temperature $\beta=1/k_B T=400$ (with $k_B=1$).
We apply linear and circularly polarized pulses of the form
\begin{align}
	&A(t) = \frac{E_0}{\Omega} \sin(\Omega (t-t_{\textrm{avg}})) \nonumber\\
	&\hspace{5mm}\times \cos^2\Big(\frac{\Omega (t-t_{\textrm{avg}}) }{2M}\Big)\cdot  \{\theta(t) - \theta(t- M\cdot T)  \}, 
\end{align}
where $E_0$ is the electric field strength, $\Omega$ the central frequency, $M$ the number of cycles, $T$ the period and $t_{\textrm{avg}}=\frac{TM}{2}$. High harmonic spectra are calculated through the formula $\abs{i\omega J_i(\omega) + \frac{\omega^2}{c} M_i(\omega) }^2$ where $J_i(\omega)$ and $M_i(\omega)$ denote the Fourier transform of the charge current $J_i(t) = N_{\boldsymbol{k}}^{-1} \sum_{\boldsymbol{k}}\textrm{Tr}[ \rho(t) v_i(\boldsymbol{k},t)]$, 
and magnetization, $M_i(t) = N_{\boldsymbol{k}}^{-1} \sum_{\boldsymbol{k}}\textrm{Tr}[ \rho(t) \sigma_i]$, in the $i=x,y,z$ direction, respectively. 
We calculate the spectra for the spin current as $\abs{i\omega J_{ij}(\omega) }^2$. Prior to the Fourier transform, we apply a Blackman window to all quantities. This is given as $f_B(t) = 0.42 - 0.5\cos(\frac{2\pi t}{M\cdot T}) +0.08\cos(\frac{4\pi t}{M\cdot T})$ \cite{Takayoshi_2019}. Care must be taken with the spin current as it has a DC component \cite{Rashba_2003}. Here, the DC-contribution at $t=0$ is subtracted before applying the Blackman window. 

\begin{figure}[t]
\centering
\includegraphics[width=\columnwidth, bb=75.6 223.20000000000002 536.4 568.8]{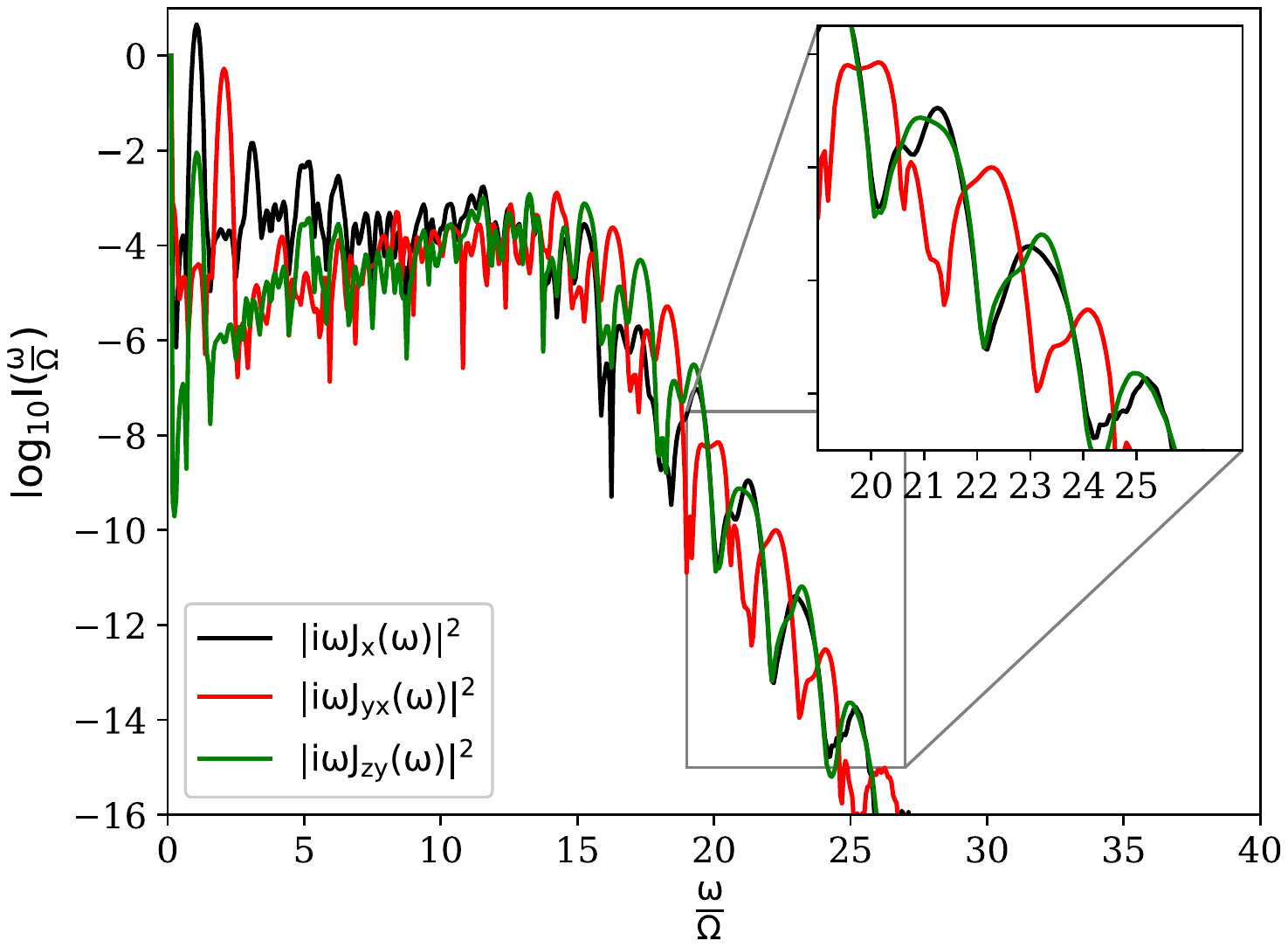}
\vspace{-10mm}
\caption{High harmonic spectra for the charge current, $J_x$, and two components of the spin current, $J_{yx}$ and $J_{zy}$, of the Rashba model with $\alpha=2.5$, $\gamma=0$, $\mu=3.5$ and $B=0$. A linearly polarized light field with $E_0=0.2$, central frequency $\Omega=0.3$ and polarization along the $x$-direction is used.
}   
\label{selectionRules}
\end{figure}

Figure \ref{selectionRules} displays the numerically obtained HHG-spectra for two different components of the spin current, $J_{yx}$ and $J_{zy}$ as well as the charge current $J_x$. To the right of the plateau, we see as expected that both $J_x$ and $J_{zy}$ display odd order harmonics, whereas $J_{yx}$ displays even order harmonics. 

\begin{figure}[t]
\centering
\includegraphics[width=\columnwidth, bb=75.6 223.20000000000002 536.4 568.8]{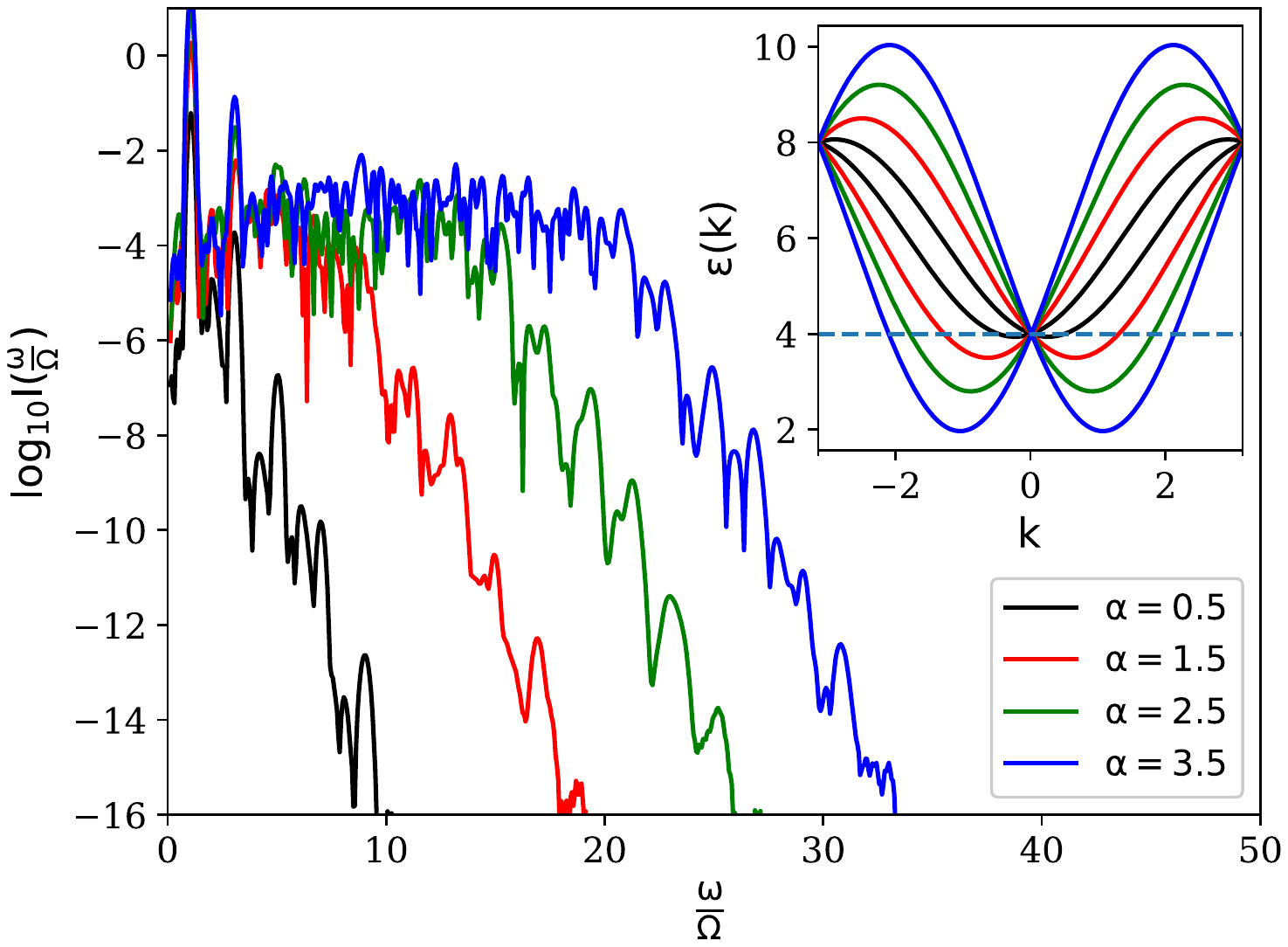}
\vspace{-10mm}
\caption{$\alpha$-dependence of the HHG spectrum for $\gamma=B=0$ and $\mu=4$. A linearly polarized light field with $E_0=0.2$, central frequency $\Omega=0.3$ and polarization along the $x$ direction is used. The inset shows the two branches of the band dispersion of Eq.~\eqref{momHam} for $k_y=0$. }   
\label{cutOff1}
\end{figure}

\begin{figure}[t]
\centering
\includegraphics[width=\columnwidth, bb=75.6 223.20000000000002 536.4 568.8]{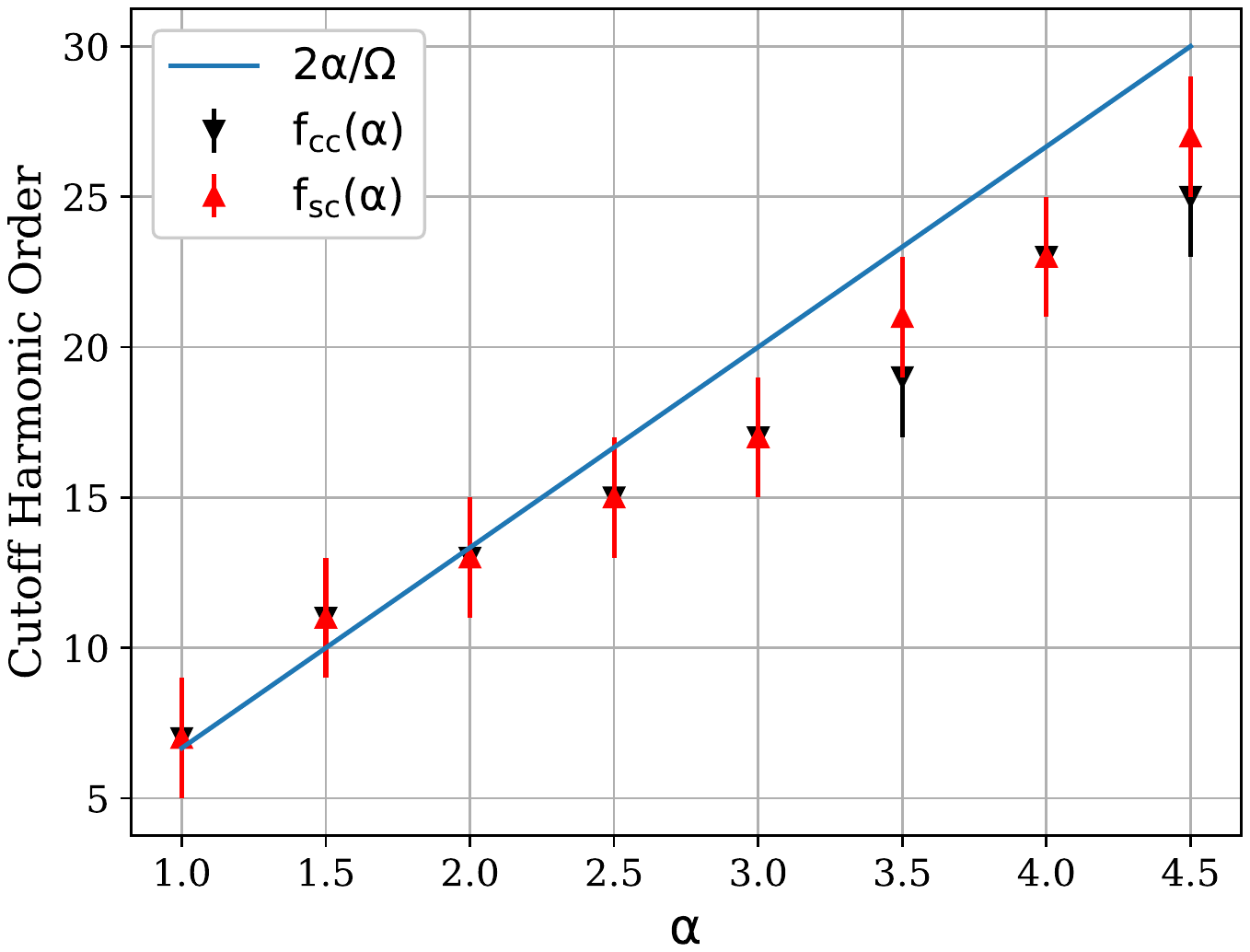}
\vspace{-10mm}
\caption{HHG cutoffs for the charge current, $f_{cc}(\alpha)$, and the $J_{zy}$ spin current component, $f_{sc}(\alpha)$ when $\mu=4$ and $\gamma=B=0$. $2\alpha/\Omega$ corresponds to the maximum band gap in units of $\Omega$. The uncertainty in determining the cutoff values is estmated to be $\delta f \approx 2.0$.}   
\label{cutOffMeasurement}
\end{figure}

{\it HHG cutoffs.} We will next demonstrate how the energy scale related to SOC, $\alpha$, manifests itself in the HHG spectra. (Swapping $\alpha$ and $\gamma$ does not produce any change.) A linearly polarized pulse is applied along the $x$-direction and the chemical potential is set to $\mu=\epsilon(\boldsymbol{k}=\boldsymbol{0})$. As shown in Fig.~\ref{cutOff1}, for $\alpha>t_h=1$, a plateau emerges, which increases with increasing $\alpha$. Upon diagonalizing Eq.~\eqref{momHam} and setting $k_y=0$, we see that the maximum energy difference between the spin split bands is $\Delta E = 2\alpha$ for $B=0$ (see inset). In units of $\Omega$, this cutoff prediction is consistent with the plateaus in Fig.~\ref{cutOff1}. In Fig.~\ref{cutOffMeasurement} we compare $\Delta E/\Omega$ to the measured cutoffs in different simulations.

Whereas in Fig.~\ref{cutOff1} the filling changes as we increase $\alpha$ while keeping $\mu$ constant, the dependence on $\alpha$ with constant filling should also be investigated. 
In the supplementary material (SM), we present spectra for simulations where the filling is fixed to the value corresponding to $\alpha=1.5$ and $\mu=4$. Since these cutoff scalings are very similar, one may conclude that the HHG cutoffs are controlled by the spin-orbit parameters rather than the filling. 

We have also measured the cutoffs in the HHG spectra for the $J_{zy}$ component of the spin current \eqref{spinCurrDef}, which closely follows the charge current (see also Fig.~\ref{selectionRules}). 
The corresponding cutoff values, $f_{sc}(\alpha)$, are presented in Fig.~\ref{cutOffMeasurement} alongside those  for the charge current, $f_{cc}(\alpha)$, and exhibit the same $\alpha$-dependence. 
Note that because there is no transverse charge current, we have a pure spin current - in line with the intrinsic spin Hall effect \cite{Sinova_2004, Shen_2004}.

\begin{figure}[t]
\centering
\includegraphics[width=\columnwidth, bb=75.6 223.20000000000002 536.4 568.8]{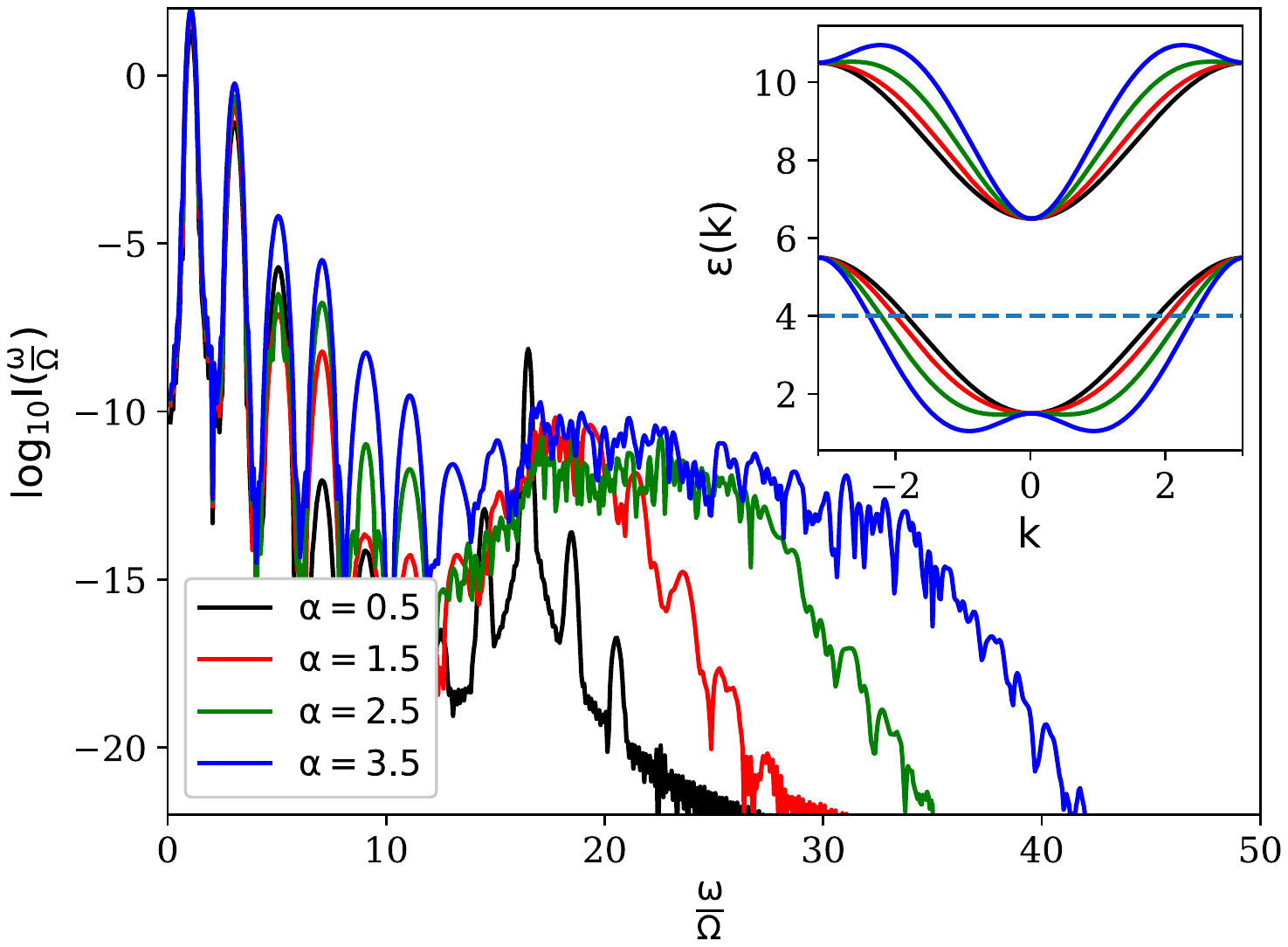}
\vspace{-10mm}
\caption{$\alpha$-dependence of the high harmonic spectrum of the Rashba model with $\gamma=0.0$, $B=2.5$ and $\mu=4$. A linearly polarized light field with $E_0=0.2$, central frequency $\Omega=0.3$ and polarization along the $x$-direction is used. }   
\label{spinPolHHG}
\end{figure}

{\it Magnetic field effects.}
Setting $B\neq 0$ will turn model \eqref{momHam} into a two-band model in the basis of eigenstates of $\hat{S}_z$. Thus, for positive $B$, the lower (upper) band will be polarized in the spin down (up) direction. Although the magnetic field/exchange field strengths considered in this section might seem high, we direct attention to a previous work where exchange fields of comparable strengths 
have been used to describe aspects of the anomalous Hall effect in a 2D Rashba ferromagnet \cite{Ado_2016}. SOC introduces a momentum-dependent inter-band matrix element vanishing at the $\Gamma$ point as well as the edges of the Brillouin zone. The result is a harmonic spectrum as shown in Fig.~\ref{spinPolHHG}. The low order harmonics show the characteristic signature of intraband harmonics, while the grouping of harmonics starting at $\omega/\Omega > 13$ can be explained by multiphoton processes across the band gap created by $B$. Indeed, the minimal band gap is $\Delta E = 2 B$ so that the minimal number of photons is $2B/\Omega \approx 13$. 
The maximum band gap is 
$2\sqrt{2\alpha^2 + B^2}$, which nicely explains the upper edges of the harmonic groupings in Fig.~\ref{spinPolHHG} (in units of $\Omega$). In contrast to previous HHG studies of two-band semiconductors \cite{Golde_2008, Wu_2015}, the high-energy part of the spectrum does not exhibit a plateau structure, but rather a dome shape. 
We interpret this as a result of the vanishing interband coupling at the $\Gamma$ point and at the Brillouin zone boundary. 

\begin{figure}[t]
\centering
\includegraphics[width=\columnwidth, bb=75.6 223.20000000000002 536.4 568.8]{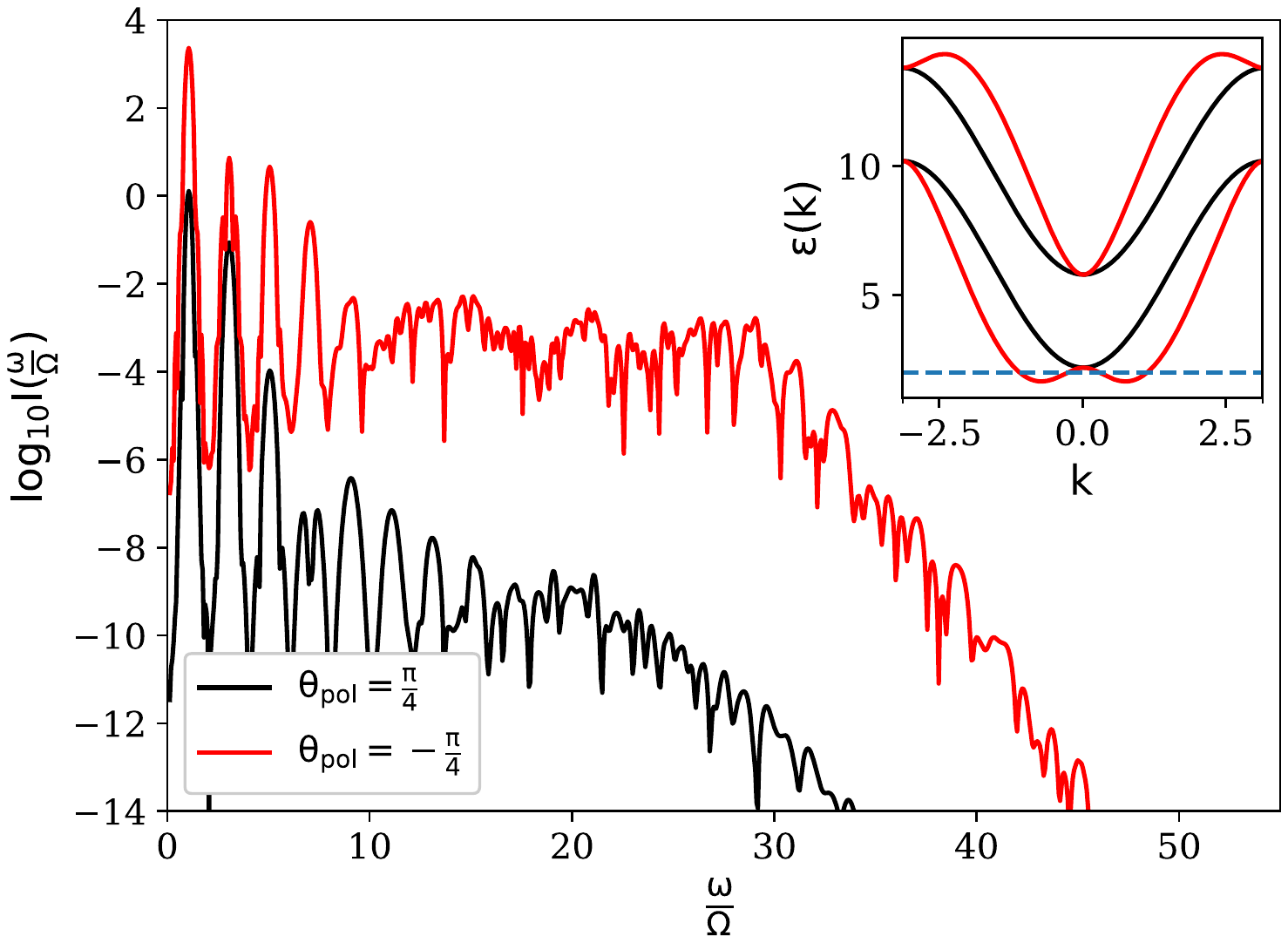}
\vspace{-10mm}
\caption{Dependence of the HHG spectra on the polarization direction. The inset shows the band structure along cuts parametrized by $k\langle 1,1\rangle$ (black) and $k\langle 1,-1\rangle$ (red), i.e., along the direction of the field polarization. 
$\alpha=\gamma=1.5$ and $B=1.8$. The field strength is set to $E_0=0.5$ and $\mu=2$.}   
\label{polDirection}
\end{figure}

If both $\alpha$ and $\gamma$ are nonzero, the Fermi surface of model (\ref{momHam}) has a nontrivial shape \cite{Manchon_2015} and it is thus interesting to ask if the magnetic-field and angular dependence of the HHG spectra allows to extract the spin-orbit parameters. 
For $\alpha=\gamma$ the energy gap is bounded by $2B \leq \Delta E \leq 2\sqrt{B^2 + 8\alpha^2}$, 
and we expect to see a difference in the spectra
when setting the linear polarization of the fields to $\Theta_{\textrm{pol}}=\pm\frac{\pi}{4}$, while measuring along the $x$ direction. The results of such calculations are shown in Fig.~\ref{polDirection}. A strong enhancement of harmonic intensity within the predicted plateau region is seen for $\Theta_{\textrm{pol}}=-\frac{\pi}{4}$ relative to $\Theta_{\textrm{pol}}=\frac{\pi}{4}$. In the SM we provide numerical evidence for the important role played by the magnetization dynamics in this case. The directional anisotropy  
appears because the spin expectation values (in equilibrium) are constant along lines with $\Theta_{\textrm{pol}}=\frac{\pi}{4}$ while they vary along $\Theta_{\textrm{pol}}=-\frac{\pi}{4}$ when $\alpha=\gamma$ \cite{Liu_2006}.

{\it Conclusions.} 
We have explored ways of extracting SOC parameters from HHG spectra. If only a Rashba or Dresselhaus coupling is present, the coupling strength can be directly deduced from the cutoff of the HHG plateau or a characteristic grouping of harmonics in strong magnetic/exchange fields. If both couplings are nonzero, insight into the relative size of the SOC parameters can be gained by studying the polarization dependence. In particular, a large change in the HHG intensity upon rotation by 90$^\circ$ indicates that $\alpha$ and $\gamma$ are of comparable magnitude. The general symmetry analysis for linearly and circularly polarized fields can help to determine relevant aspects of a microscopic model on the basis of HHG spectra, at least for systems with strong SOC. We have also shown that the $J_{zy}$ spin current is strongly correlated with the $J_x$ charge current and that both follow the same cutoff scaling with increasing $\alpha$. Since there is much interest in the control of spin currents, high harmonic generation and detection methods may be useful for identifying SOC materials with ideal properties for spintronics applications.

{\it Acknowledgments}
ML and PW acknowledge support from ERC Consolidator Grant No. 724103. YM acknowledges support from Grant-in-Aid for Scientific Research from JSPS, KAKENHI Grant Nos. JP19K23425, JP20K14412, JP20H05265, and JST CREST Grant No. JPMJCR1901. M. S. acknowledges financial support from the U. S. Department of Energy (DOE), Office of Basic Energy Sciences, Division of Materials Sciences and Engineering, under contract no. DE-AC02-76SF00515, and Alexander von Humboldt Foundation for its support with a Feodor Lynen scholarship.

\bibliography{bibliography}

\clearpage

\widetext

\section{High-harmonic generation in spin-orbit coupled systems - Supplementary material}

\section{Selection rules -- a simple example}

To illustrate the formalism introduced in the paper on a simple example, we derive the well known result that inversion symmetry implies only odd order harmonics. We consider the Hamiltonian 
\begin{equation}
	\hat{H}(t) = \sum_{{k}} \cos({k} + A_0 \cos(\Omega t)) \hat{c}^{\dagger}_{{k}} \hat{c}_{{k}},
\end{equation}
and, since we are interested in HHG, choose as operator $\mathcal O$ the charge velocity 
\begin{equation}
	v({k},t) = \frac{\partial}{\partial {k}}h({{k}},t) =- \sin({k} + A_0 \cos(\Omega t)),
\end{equation}
which yields the current $J(t) = \sum_{{k}} v({k},t) \langle\hat{c}^{\dagger}_{{k}}\hat{c}_{{k}}\rangle$. Whereas $h({k},t) =  h(-{k},t)$ in equilibrium ($A_0=0$), in the presence of the drive, we need to extend the symmetry operation to include time as follows
\begin{equation}
  \mathcal{P} \otimes \mathcal{T}_2 \otimes \boldsymbol{1} \equiv \begin{cases}
	&{k} \rightarrow -{k} \nonumber\\
	&t \rightarrow t + T/2 \\
  \end{cases}.
\end{equation}
Clearly, the group generated by this operation is isomorphic to $\boldsymbol{Z}_2$ if we identify $t$ with $t+T$. Labelling the group element above as $g$, we expand both sides of
\begin{equation}
	 \hat{g} {v}({k},t) e^{in\Omega t} \hat{g}^{-1} = {v}({k},t) e^{in\Omega t} 
\end{equation}
to obtain
\begin{equation}
	v(-{k},t + T/2) e^{in\Omega (t + T/2)} = v({k},t) e^{in\Omega t}. 
\end{equation}
Since $v(-{k},t + T/2)= -v({k},t)$, $n$ is constrained by $e^{in\pi}=-1$, which implies that $n$ is odd. 

\section{Consequences of the selection rules}

The Hamiltonian for which we will study selection rules is once again written as 
\begin{align}
    &\hat{H} = \sum_{\boldsymbol{k}}  \Psi_{\boldsymbol{k}}^{\dagger} [\epsilon(\boldsymbol{k})\otimes \sigma_0 -(\alpha \sin(k_y a) -\gamma \sin(k_x a))\otimes \sigma_x \nonumber\\
    &\quad \quad +(\alpha \sin(k_x a) -\gamma \sin(k_y a)) \otimes \sigma_y + B \sigma_z] \Psi_{\boldsymbol{k}} ,
    \label{momHam}
\end{align}
with $\epsilon(\boldsymbol{k})=2t_h(4-\cos(k_x a) - \cos(k_y a))$. The band dispersions are 
\begin{equation} \label{bandDisp}
	\epsilon_{\pm}(\boldsymbol{k}) = \epsilon(\boldsymbol{k}) \pm \sqrt{ (\alpha \sin(k_y a) -\gamma \sin(k_x a))^2 + (\alpha \sin(k_x) -\gamma \sin(k_y))^2 + B^2 }.
\end{equation}
In the following we set $t_h=a=1$.

{\it Linearly polarized light.}
For linearly polarized light described by $A_x(t) = A_0 \cos(\Omega t)$, we consider the charge velocity
\begin{equation} \label{chargeVelR}
\begin{aligned}
	v_x(\boldsymbol{k},t) =&\frac{\partial }{\partial k_x}h(\boldsymbol{k},t) 
	= 2 \sin(k_x + A_x(t)) \cdot \boldsymbol{1}
	+ \alpha \cos(k_x + A_x(t)) \cdot \sigma_y 
	 + \gamma \cos(k_x + A_x(t)) \cdot \sigma_x \\
\end{aligned}
\end{equation}
and the symmetry ${g}\equiv\mathcal{P} \otimes \mathcal{T}_2 \otimes \mathcal{S}_{(-)}$. Since the spin transformation does not mix the $\boldsymbol{1}$, $\sigma_x$ and $\sigma_y$ matrices it is sufficient to consider
\begin{equation}
\begin{aligned}
	& \hat{g}\cos(k_x + A_x(t))e^{in \Omega t} \sigma_{x,y}\hat{g}^{-1}
	=\cos(-k_x - A_x(t))e^{in \Omega (t+T/2)} (-\sigma_{x,y}) 
	\stackrel{!}{=}\cos(k_x + A_x(t))e^{in \Omega t} \sigma_{x,y}, \\
\end{aligned}
\end{equation}
which leads to $e^{in\pi}=-1$, i.e., odd order harmonics only. 
The non-trivial relation between spin and momentum in spin-orbit coupled systems gives rise to interesting spin dynamics upon radiation. In linearly polarized light, we consider $ \mathcal{P} \otimes \mathcal{T}_2 \otimes \mathcal{S}_{(-)}$. That the spin operators
in the $x$ and $y$ direction have \emph{odd} harmonics can be easily seen as follows
\begin{equation}
	g\sigma_{x,y}e^{in\Omega t} {g}^{-1} = -\sigma_{x,y}e^{in\Omega (t + T/2)}\stackrel{!}{=} \sigma_{x,y}e^{in\Omega t},
\end{equation}
whereas for the $z$-component, we have
\begin{equation}
	{g}\sigma_{z}e^{in\Omega t} {g}^{-1} = \sigma_{z}e^{in\Omega (t + T/2)} \stackrel{!}{=} \sigma_{z}e^{in\Omega t}, 
\end{equation}
which implies \emph{even} harmonics. An oscillating magnetization will on top of any charge current contribute to the power spectrum through $I_{i}(\omega)\equiv \abs{i\omega J_i(\omega)  + (\frac{\omega^2}{c})M_i(\omega)}^2$ with $M_i(\omega)=N_{\boldsymbol{k}}^{-1}\sum_{\boldsymbol{k}}\expval{\sigma_i}(\omega)$, where $c$ is the speed of light and $N_{\boldsymbol{k}}$ is the number of ${\boldsymbol{k}}$ points. While the expectation value of $\sigma_z$ is zero in equilibrium states of the Rashba model, circularly polarized light could induce a nonzero expectation value and hence even harmonics - specifically $n=0,4,8,12,...$ from Eqs. \eqref{circSymClockwise} and \eqref{circSymAntiClockwise} below \cite{Zhou_2007}.

{\it Circularly polarized light.}
To consider the effect of circularly polarized light, we begin by noting that 
\begin{equation} \label{chargeVelx}
\begin{aligned}
	v_x({\boldsymbol{k}},t) &=  2 \sin(k_x + A_x(t)) \cdot \sigma_0
	+ \alpha \cos(k_x + A_x(t)) \cdot \sigma_y 
	+ \gamma \cos(k_x + A_x(t)) \cdot \sigma_x \\
\end{aligned}
\end{equation}
and
\begin{equation} \label{chargeVely}
\begin{aligned}
	v_y({\boldsymbol{k}},t) &=  2 \sin(k_y + A_y(t)) \cdot \sigma_0
	- \alpha \cos(k_y + A_y(t)) \cdot \sigma_x 
         - \gamma \cos(k_y + A_y(t)) \cdot \sigma_y .
\end{aligned}
\end{equation}

\subsection{Rashba model}
To investigate the selection rules for circularly polarized harmonics, we define the following symmetry
\begin{equation} \label{circSymClockwise}
  \mathcal{R}_{90^{\circ}} \otimes \mathcal{T}_4 \otimes \mathcal{S}_{90^{\circ}} =\begin{cases}
	& (k_x,k_y) \rightarrow (k_y,-k_x) \\
	&t \rightarrow t + T/4 \\	
	&  (\sigma_x, \sigma_y, \sigma_z) \rightarrow (\sigma_y,-\sigma_x, \sigma_z) 
  \end{cases}
\end{equation}
(also given in the paper) which is a symmetry valid for 
\begin{align}
	A_x &= A_0 \sin(\Omega t), \quad A_y = A_0 \cos(\Omega t) 
	\label{circLight}
\end{align}
and $\gamma=0$. Note that the time translation $t \rightarrow t + T/4$ results in the same type of rotation for the vector potential as in the spatial sector and the spin sector, namely 
$(A_x(t),A_y(t)) \rightarrow (A_y(t),-A_x(t))$. 
The invariance of $\hat H$ follows from 
\begin{equation}
\begin{aligned}
	&\hspace{-1cm}[ \mathcal{R}_{90^{\circ}} \otimes \mathcal{T}_4 \otimes \mathcal{S}_{90^{\circ}}] {h(\boldsymbol{k}, t)} [\mathcal{R}_{90^{\circ}} \otimes \mathcal{T}_4 \otimes \mathcal{S}_{90^{\circ}}]^{-1} \\
	=&- (\alpha \sin(-k_x - A_x(t)) - \gamma \sin(k_y + A_y(t))) \otimes \sigma_y \\
	&+ (\alpha \sin(k_y + A_y(t)) - \gamma \sin(-k_x - A_x(t))) \otimes (-\sigma_x) \\
	\stackrel{!}{=}& - (\alpha \sin(k_y + A_y(t)) - \gamma \sin(k_x + A_x(t))) \otimes \sigma_x \\
	&+ (\alpha \sin(k_x + A_x(t)) - \gamma \sin(k_y + A_y(t))) \otimes \sigma_y  = {h(\boldsymbol{k}, t)} \\
\end{aligned}
\end{equation}
implying that $\gamma$ must be zero. Hence, we have found a symmetry of the Rashba model in circularly polarized light. The quantity we are interested in is the emitted radiation with circular polarization for which we define the charge velocities 
\begin{equation}
	v_{\pm}(k,t) = v_x(k,t) \pm i v_y(k,t) ,
\end{equation}
where +(-) refers to right and left hand circular polarization, respectively. We have for $\gamma=0$
\begin{equation}
\begin{aligned}
	v_{\pm}(k,t) &= 2\big( \sin(k_x + A_x(t)) \pm i \sin(k_y + A_y(t)) \big) \\
	&+\alpha \cos(k_x + A_x(t)) \sigma_y \mp i \alpha \cos(k_y + A_y(t)) \sigma_x \\
\end{aligned}
\end{equation}
and
\begin{equation}
\begin{aligned}
	&[ \mathcal{R}_{90^{\circ}} \otimes \mathcal{T}_4 \otimes \mathcal{S}_{90^{\circ}}] v_{\pm}(k,t) e^{in\Omega t}[ \mathcal{R}_{90^{\circ}} \otimes \mathcal{T}_4 \otimes \mathcal{S}_{90^{\circ}}]^{-1} \\
	&= e^{in\Omega(t + T/4)}\{2\big( \sin(k_y + A_y(t)) \pm i \sin(-k_x - A_x(t)) \big) \\
	&+\alpha \cos(k_y + A_y(t)) (-\sigma_x) \mp i \alpha \cos(-k_x - A_x(t)) \sigma_y  \} \\
	&\stackrel{!}{=}v_{\pm}(k,t) e^{in\Omega t}, \\
\end{aligned}
\end{equation}
from where we obtain the following requirement:
\begin{equation}
	e^{in\frac{\pi}{2}} = \pm i . \label{selection_circular}
\end{equation}

The symmetries just derived also hold for the model without SOC terms, which has the same selection rules. As detailed in Ref.~\onlinecite{Neufeld_2017}, the power spectrum can be calculated as $I_{\pm}(n\Omega) = \abs{a_{\pm}(n\Omega)}^2$, with 
\begin{equation}\label{circPolFormula}
	a_{\pm}(n\Omega) = \mathcal{FT}\Big[\frac{d}{dt}J_{x}(k,t) \pm i \frac{d}{dt}J_{y}(k,t) \Big].
\end{equation}

\subsection{Dresselhaus model}
For the Dresselhaus model, we have a different symmetry
\begin{equation}  \label{circSymAntiClockwise}
  \mathcal{R}_{90^{\circ}} \otimes \mathcal{T}_4 \otimes \mathcal{S}_{-90^{\circ}} =\begin{cases}
	& (k_x,k_y) \rightarrow (k_y,-k_x)  \\
	&t \rightarrow t + T/4 \\
	&  (\sigma_x, \sigma_y, \sigma_z) \rightarrow (-\sigma_y, \sigma_x, \sigma_z) 
  \end{cases}
\end{equation}
and a calculation completely analogous to that of the previous subsection yields $e^{in\frac{\pi}{2}} = \pm i$ as for the Rashba model.

\section{Additional Results}

\subsection{Harmonic orders}
 
\begin{figure}[ht]
  {
	\begin{minipage}[c][1\width]{
	   0.3\textwidth}
	   \centering
	   \includegraphics[width=\textwidth]{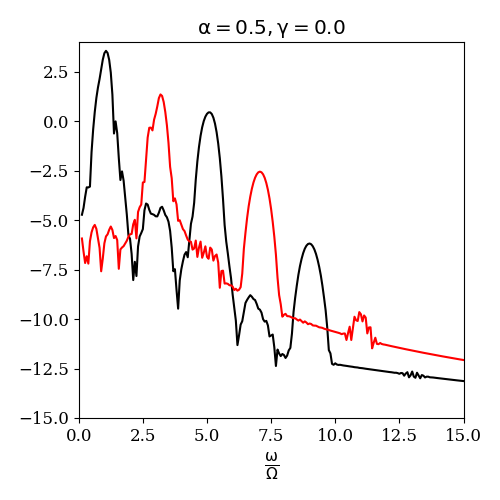}
	\end{minipage}}
 \hfill 	
  {
	\begin{minipage}[c][1\width]{
	   0.3\textwidth}
	   \centering
	   \includegraphics[width=\textwidth]{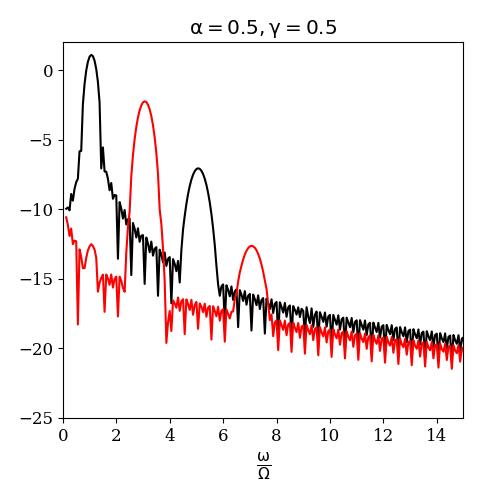}
	\end{minipage}}
 \hfill	
  {
	\begin{minipage}[c][1\width]{
	   0.3\textwidth}
	   \centering
	   \includegraphics[width=\textwidth]{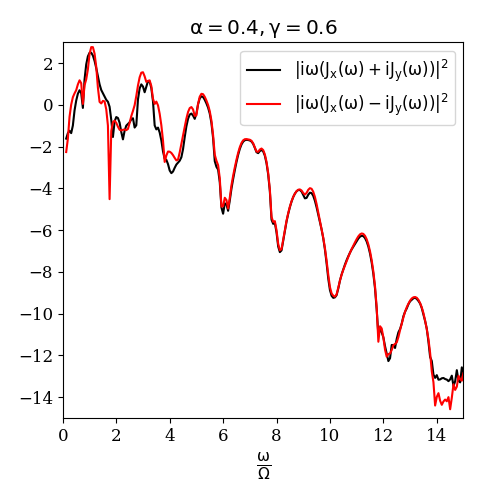}
	\end{minipage}}
\caption{High harmonic spectra for three sets of $(\alpha,\gamma)$ where $\mu=5$ and $E_0=0.2$ and right hand circularly polarized light is applied.}
\label{circPolFig}
\end{figure}

To help with the interpretation of this section, we remind the reader of a symmetry which we expect to hold for both circularly and linearly polarized light
\begin{equation}
  \mathcal{P} \otimes \mathcal{T}_2 \otimes \mathcal{S}_{(-)} =\begin{cases}
	&\boldsymbol{k} \rightarrow -\boldsymbol{k} \nonumber\\
	&t \rightarrow t + T/2 \\
	& (\sigma_x, \sigma_y, \sigma_z) \rightarrow (-\sigma_x,-\sigma_y, \sigma_z) \nonumber
  \end{cases}.
\end{equation}
The presented selection rule is however not the entire story when it comes to the Rashba Dresselhaus model with $\alpha=\gamma$,
\begin{equation}
	h(\boldsymbol{k}) = \epsilon(\boldsymbol{k})\sigma_0 - \alpha (\sigma_x \pm \sigma_y) (\sin(k_y) \mp \sin(k_x)).
\end{equation} 
To illustrate this, we present results for circular polarization for three different sets of $(\alpha, \gamma)$ in Fig.~\ref{circPolFig}. HHG intensities for right and left hand polarized harmonics are calculated as in Eq.~\eqref{circPolFormula}. The leftmost panel illustrates that the selection rules 
following from Eq.~(\ref{selection_circular}) 
hold. When $\alpha=\pm \gamma$, we find the following symmetry
\begin{equation}
  \mathcal{P} \otimes \mathcal{T}_2 \otimes \mathcal{S}_{x,\mp y} =\begin{cases}
	&\boldsymbol{k} \rightarrow -\boldsymbol{k} \nonumber\\
	&t \rightarrow t + T/2 \\
	& \sigma_{x(y)} \rightarrow \mp \sigma_{y(x)} \nonumber
  \end{cases},
\end{equation}
and at the same time no symmetry involving $t \rightarrow t + T/4$ from which we would anticipate only odd order harmonics (and not the pattern seen in the leftmost panel). The observation that the middle panel still bears signatures of such a symmetry is explained by the fact that $(\sigma_x \pm \sigma_y)$ is a constant of motion. Lastly, the rightmost spectrum bears the characteristic of a system with a two-fold rotational symmetry and harmonics of all odd orders are allowed for both chiralities of the emitted radiation.

\subsection{Role of filling and chemical potential}

\begin{figure}[ht]
  {
	\begin{minipage}[c][1\width]{
	   0.47\textwidth}
	   \centering
	   \includegraphics[width=\textwidth]{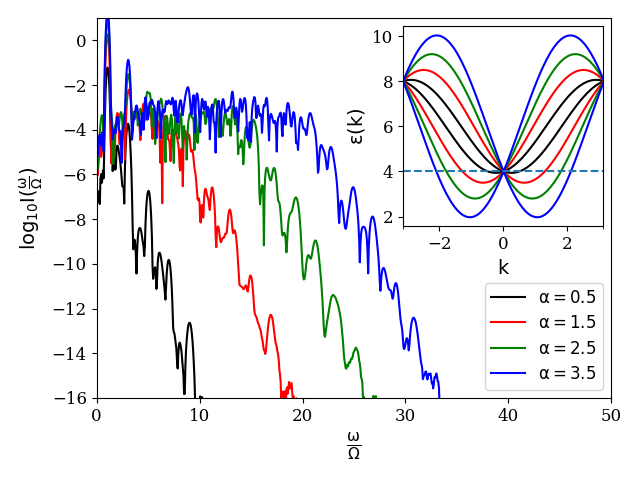}
	\end{minipage}}
 \hfill 	
  {
	\begin{minipage}[c][1\width]{
	   0.47\textwidth}
	   \centering
	   \includegraphics[width=\textwidth]{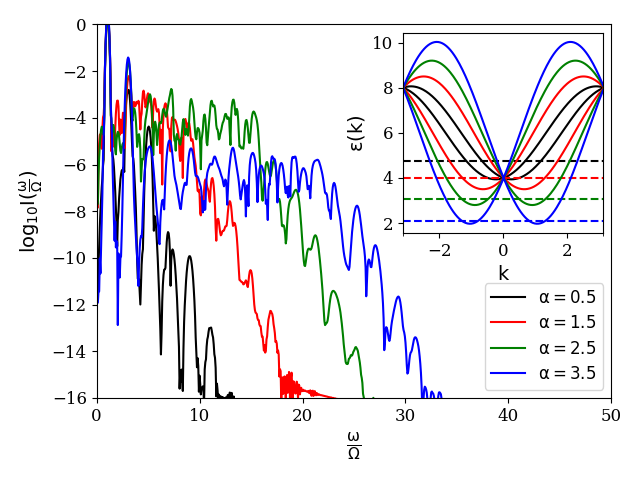}
	\end{minipage}}
\caption{Figure showing the influence of keeping a constant chemical potential (left panel) and of keeping the filling constant while changing the chemical potential (right). }
\label{fillingSupFig}
\end{figure}

In Fig.~\ref{fillingSupFig}, we present two panels where we compare parameter sweeps of $\alpha$ for constant $\mu$ (left panel) and constant filling (right panel). While the intensities of the HHG spectra are different, the cutoff scaling with $\alpha$ is the same in both cases.

\subsection{Effect on the magnetization}
\begin{figure}[ht]
  {
	\begin{minipage}[c][1\width]{
	   0.47\textwidth}
	   \centering
	   \includegraphics[width=\textwidth]{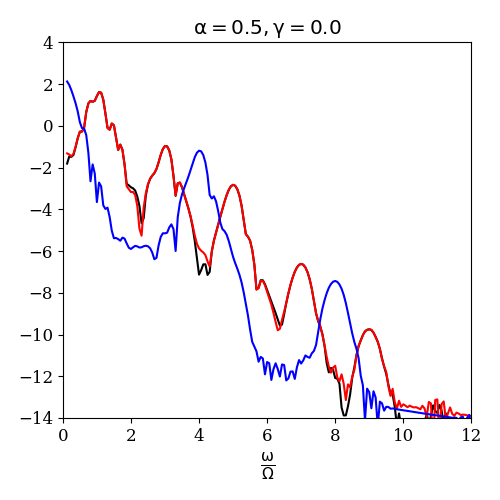}
	\end{minipage}}
 \hfill 	
  {
	\begin{minipage}[c][1\width]{
	   0.47\textwidth}
	   \centering
	   \includegraphics[width=\textwidth]{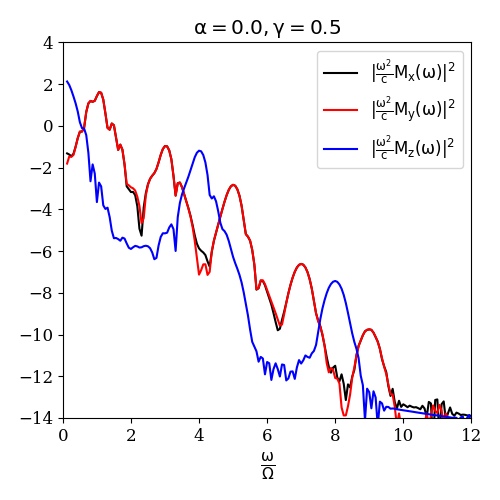}
	\end{minipage}}
\caption{Figure displaying the harmonic spectra for the magnetizations only. The driving field strength is $E_0=0.2$ and $\mu=4.0$.}
\label{magCircFig}
\end{figure}

We find that the ratio $\abs{(\frac{\omega^2}{c}) M(\omega)} /\abs{i\omega J(\omega)}$ is small, and that the total spectrum is not qualitatively changed by it. The following results should thus be regarded as a validation of the selection rules in circularly polarized light presented in the paper and not as an experimentally detectable spectrum. Figure~\ref{magCircFig} shows $\abs{(\frac{\omega^2}{c}) M_i(\omega)}^2$ for $i=x,y,z$ where the left and right panels correspond to $\alpha=0.5, \gamma=0.0$ and $\alpha=0.0, \gamma=0.5$ respectively. These figures should be interpreted in light of the selection rules following from $\mathcal{R}_{90^{\circ}} \otimes \mathcal{T}_4 \otimes \mathcal{S}_{90^{\circ}}$ in equation \eqref{circSymClockwise}, relevant for the left panel, and $\mathcal{R}_{90^{\circ}} \otimes \mathcal{T}_4 \otimes \mathcal{S}_{-90^{\circ}}$, Eq.~\eqref{circSymAntiClockwise} for the right panel. The observed harmonics are compatible with the selection rules following from the symmetries just presented.

\subsection{Anisotropy}
\begin{figure}[h]
  \centering
  \includegraphics[angle=-0, width=\textwidth]{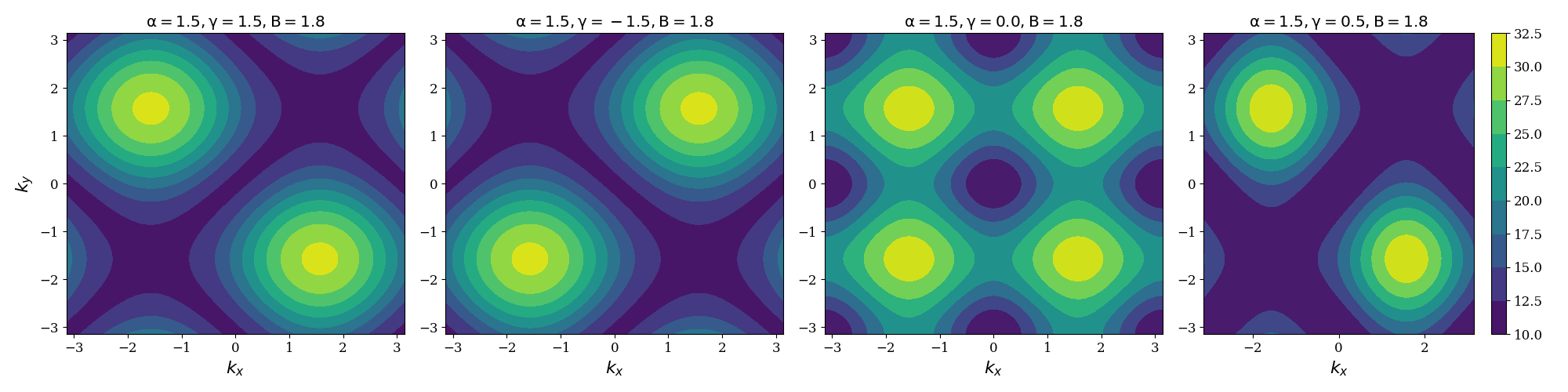}
  \caption{Plot of the band gaps for various models according to equation \eqref{bandGapEquation}. }
  \label{Band_gaps}
\end{figure}

Figure~\ref{Band_gaps} aims to clarify why for certain choices of $\alpha$ and $\gamma$ the harmonic intensity may be enhanced in one direction compared to the orthogonal direction, as shown in Fig.~5 in the paper. To this end, we plot the band gap, which is given by 
\begin{equation} \label{bandGapEquation}
	\Delta E = 2\sqrt{ (\alpha \sin(k_y a) -\gamma \sin(k_x a))^2 + (\alpha \sin(k_x) -\gamma \sin(k_y))^2 + B^2 }
\end{equation}
in the Brillouin zone. 
In the two leftmost panels of Fig.~\ref{Band_gaps}, the four-fold rotational symmetry implicit in some of the symmetries in the paper is clearly broken. Restricting attention to the leftmost panel, a field acceleration along $\theta_{\textrm{pol}}=-\frac{\pi}{4}$ will give rise to enhanced intensity relative to the orthogonal direction because according to how the expectation value of the spin changes across the Brillouin zone, this direction will give rise to magnetization dynamics, whereas this happens to a much lesser extent along the $\theta_{\textrm{pol}}=\frac{\pi}{4}$ direction (see Fig.~2 in Ref.~\onlinecite{Liu_2006}). While the magnetic dipole contribution to the radiation may be small, the Pauli spin operators  enter into the expression for the charge current - a consequence of spin-momentum locking. (See for instance Eqs.~\eqref{chargeVelx} and \eqref{chargeVely}.) Thus, their dynamics will strongly affect the resulting HHG spectrum. Also the different band structures for cuts along $\theta_{\textrm{pol}}=\pm\frac{\pi}{4}$ through the $\Gamma$ point suggest an effect of the polarization direction on the HHG spectrum. Note that this anisotropy is absent if $\gamma=0$ (or $\alpha=0$), and indeed there is no polarization dependence in the HHG spectrum in this case.

\begin{figure}[ht]
  {
	\begin{minipage}[c][1\width]{
	   0.47\textwidth}
	   \centering
	   \includegraphics[width=\textwidth]{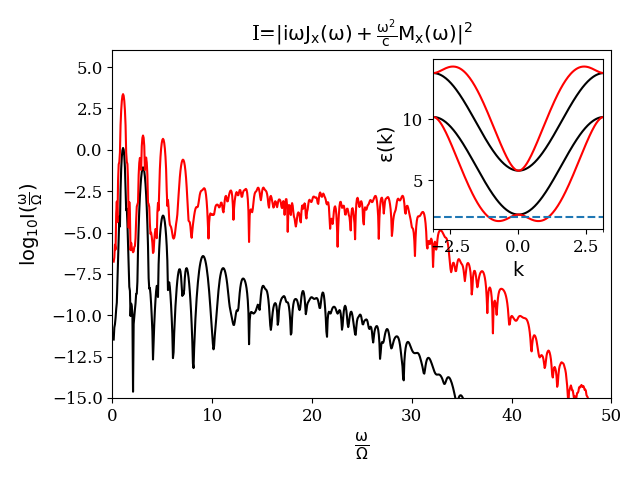}
	\end{minipage}}
 \hfill 	
  {
	\begin{minipage}[c][1\width]{
	   0.47\textwidth}
	   \centering
	   \includegraphics[width=\textwidth]{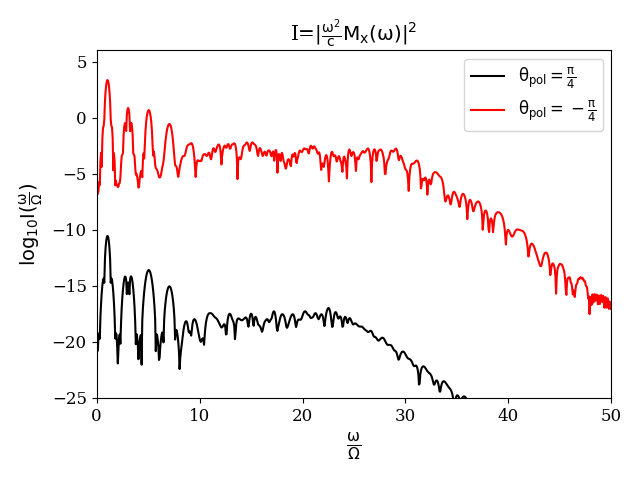}
	\end{minipage}}
\caption{Figure showing the anisotropy for the radiated intensity in the left panel and the magnetization $M_x$ in the right panel. The field strength is $E_0=0.5$ here.}
\label{magSupFig}
\end{figure}

\begin{figure}[h]
  \centering
  \includegraphics[angle=-0, width=0.5\textwidth]{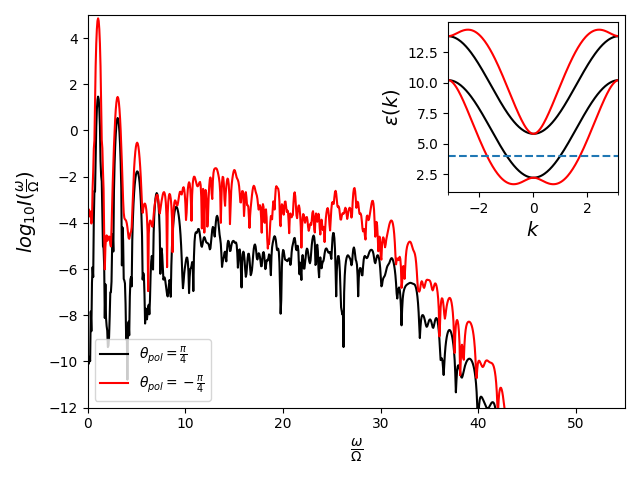}
  \caption{Figure showing the anisotropy for radiated intensity with parameters as in Fig. \ref{magSupFig} but with $\mu=4$. }
  \label{mu4polDir}
\end{figure}

 In Fig.~\ref{magSupFig}, two panels corresponding to the charge current (left panel) and the magnetization scaled by a $\big(\frac{\omega^2}{c}\big)$ factor (right panel) are shown. The right panel indicates the importance played by magnetization dynamics in generating the resulting spectra seen in the left panel. Further evidence of the role played by the dynamics of $\sigma_x, \sigma_y$ in the charge current is found by noting the similarity in the cutoff positions in both panels. In addition to this, both for $\theta_{\textrm{pol}}=\pm\frac{\pi}{4}$, the upper bound of $\Delta E$ coincides well with the observed cutoffs. Lastly, to show that this is generic for the given Hamiltonian, we present a simulation with $\mu=4$ in Fig.~\ref{mu4polDir}.

\end{document}